\documentstyle{article}
\font\tenbf=cmbx10
\font\tenrm=cmr10
\font\tenit=cmti10
\font\elevenbf=cmbx10 scaled\magstep 1
\font\elevenrm=cmr10 scaled\magstep 1
\font\elevenit=cmti10 scaled\magstep 1

\font\ninerm=cmr9

\def\ETslash{\not{\hbox{\kern-4pt $E_T$}}}

\def\ra{\rightarrow}

\def\width{ \Gamma( t \ra b W^+) }
\def\jourtpol#1&#2{{\it #1}{\bf #2}}
\def\journal#1&#2(#3)#4{
{\unskip,~\it #1\unskip~\bf\ignorespaces #2\unskip~\rm (19#3) #4}}
\def\reflist{\section*{REFERENCES\markboth
     {REFLIST}{REFLIST}}\list
     {[\arabic{enumi}]\hfill}{\settowidth\labelwidth{[999]}
     \leftmargin\labelwidth
     \advance\leftmargin\labelsep\usecounter{enumi}}}
 \relax

\begin{document}
\begin{center}
\makebox[\textwidth] [r] {MSUHEP-93/25}
\makebox[\textwidth] [r] {October 1993}
\end{center}
\begin{center}{{\tenbf REPORT OF THE SUBGROUP ON THE TOP QUARK
\footnote{\ninerm\baselineskip=11pt
To appear in the proceedings of the Workshop on Physics at Current
Accelerators and the Supercollider, Argonne, IL, 2-5 June1993.
}\\}
\vglue 1.0cm
{\tenrm  C.--P. Yuan$^{(a)}$ and L. Nodulman$^{(b)}$  \\}
\vglue 0.3cm
{\tenrm and\\}
\vglue 0.3cm
{\tenrm  H. Baer$^{(c)}$, V. Barger$^{(d)}$, D.O. Carlson$^{(a)}$,
Chih-Hao Chen$^{(c)}$,
J.L. Diaz-Cruz$^{(e)}$,  W.S. Hou$^{(f)}$,    \break
G.L. Kane$^{(g)}$, C.~Kao$^{(c)}$,
G.A. Ladinsky$^{(a)}$, C.S. Li$^{(h)}$,
A. Pomarol$^{(i)}$, M.H. Reno$^{(j)}$, C. Schmidt$^{(i)}$  \\}
\vglue 0.3cm
\baselineskip=13pt
{\tenit {\rm ($a$)} Michigan State University,
 {\rm ($b$)} Argonne National Laboratory, \break
{\rm ($c$)} Florida State University,
{\rm ($d$)} University of Wisconsin at Madison,
{\rm ($e$)} Cinvestav, Mexico, \break
{\rm ($f$)} National Taiwan University, Taiwan,
{\rm ($g$)} University of Michigan,
{\rm ($h$)} Chongqing University, China,
{\rm ($i$)} University of California at Santa Cruz,
{\rm ($j$)} University of Iowa.
}

\vglue 0.8cm
{\tenrm ABSTRACT}}
\end{center}
\vglue 0.3cm
{\rightskip=3pc \leftskip=3pc \tenrm\baselineskip=12pt \noindent
The top group studied discovery issues
as well as measurements to be made at the Tevatron,
the LHC and the SSC.
\vglue 0.6cm}

\baselineskip=14pt
\elevenrm

\def\ra{\rightarrow}
\def\it{\elevenit}


\vspace{0.4cm}
\noindent\makebox[\textwidth][l]{\bf 1. Introduction }
\vspace{0.1cm}

The top physics section was a subgroup of the heavy flavors group.
The results compiled in this report are based on
a review talk
given by C.--P. Yuan [\ref{china}],
and reports presented at both the
SSC Physics Symposium held in Madison, Wisconsin on 29--31 March 1993 and
the Workshop on Physics at Current
Accelerators and the Supercollider held in Argonne, Illinois on 2--5 June
1993. To make this report more informative, we also include
some recent results in the literature.

\vspace{0.4cm}
\noindent\makebox[\textwidth][l]{\bf 2. Why Study the Top Quark?}
\vspace{0.1cm}

The top quark is a crucial element of the Standard Model (SM).
The top quark has been found to be heavier than 45 GeV from
the SLAC and LEP experiments
and 108 (103) GeV from CDF (D0) data assuming different top production cross
sections [\ref{nmshaw},\ref{diehl}].
The first limit is model independent while the second limits are for the
Standard Model top quark. A model independent limit derived from
$W$ and $Z$ cross sections at CDF of 62 GeV was also presented [\ref{nmshaw}].
Upper bounds on $m_t$ can be obtained from examining the
radiative corrections to low energy observables, such as the $\rho$
parameter, which are proportional to $m_t^2$ at the one loop level
and thus sensitive to $m_t$.
The consistency of all the low energy experimental data
requires $m_t$ to be less than about
200 GeV.  Overall fits, dominated by LEP data, give a top mass of
about 160 GeV with a statistical error of about 20 GeV and a systematic
variation with the Higgs mass of another $\pm$20 GeV [\ref{slwu}].

Since the top quark is heavy, of the
same order of magnitude as the $W$--boson mass, any physical observable
related to the top quark may be sensitive to new physics.
The top quarks will therefore allow many new tests of the SM and new probes of
physics at the 100 GeV scale [\ref{toppol}].
The most important consequence of a heavy top quark
is that to a good approximation it decays as a free quark, since its
lifetime is short and it does not have time to
bind with light quarks before it decays [\ref{decay}].
 Thus, we can use the polarization
properties of the top quark
as an additional tool, testing the SM and
probing for new physics [\ref{toppol}].
 Furthermore, because the heavy top quark
has the weak two--body decay $t \ra b W^+$, it will
analyze its own polarization.

 First, we
discuss the production
mechanism for top quarks at hadron colliders, then we
discuss the decay modes and branching ratios
for top quarks. We also report on how well the mass and
width of the top quark can be measured, and
how to detect CP violation effects in top quarks studies.

\vspace{0.4cm}
\noindent\makebox[\textwidth][l]{\bf 3. How to Produce Top Quarks}
\vspace{0.1cm}

At the Tevatron, the dominant production mechanisms for a SM top quark are
the QCD processes $q \bar q, \ gg \ra t \bar t$ .
For a heavy top quark, $m_t > 100$ GeV, the $q \bar q$ process
becomes most important.
The full next-to-leading-order calculation for these QCD processes was
completed several of years ago [\ref{nloqcd}].
The electroweak radiative corrections to these processes were also
calculated in Refs.~{[\ref{qcdtpol}]} and {[\ref{qqttqcd}]}.
Therefore, the production rates for top quark pairs
at hadron colliders are  well predicted.

If the top quark is as heavy as 140 GeV, then another production
mechanism known as the $W$--gluon fusion process becomes
 important [\ref{wgone},\ref{wgtwo}].
The production mechanism of the latter process involves the electroweak
interaction, therefore it can probe the electroweak sector of
the theory. This is in contrast to the usual QCD production mechanism
which only probes the QCD interaction when counting the
top quark event rates.

At the SSC/LHC, the dominant production mechanism for a SM top quark is
the QCD process $gg \ra t \bar t$. The subprocess
$q \bar q \ra t \bar t$ is always small compared with the gluon--gluon
fusion process, even into the TeV region. In one year there will be about
$10^8$ $t \bar t$ pairs produced at the SSC from the QCD
processes. The $W$--gluon fusion process is also important.
For a 140 GeV top quark, about $10^7$ top quarks per year are produced via
this mechanism at the SSC. The production rates of the top quark
at the LHC, given ten times higher luminosity,
are about equal to those at the SSC.

\vfill
\eject

\noindent\makebox[\textwidth][l]{\bf 4. Polarization of Top Quarks}
\vspace{0.1cm}

In the SM, the heavy top quarks produced from the Born level
QCD processes are unpolarized.
At the one loop level, the top quarks are {\it transversely}
(perpendicular to the scattering plane)
polarized at the level of a couple of percent
[\ref{goldstein},\ref{toppol},\ref{trans}].
Top quarks will have longitudinal
polarization if weak effects are present in their production.
The polarization of the top quark produced from the usual QCD process
after including the electroweak radiative
corrections was discussed in Ref.~{[\ref{qqttqcd}]}.
In the SM, the heavy top quark produced via the $W$--gluon
fusion process is left--handed polarized.
If new interactions occur, they may manifest themselves in
an enhancement of the polarization effects in
the production of the top
quark via the $W$--gluon fusion process [\ref{toppol}].

\vspace{0.4cm}
\noindent\makebox[\textwidth][l]{\bf 5. How Top Quarks Decay}
\vspace{0.1cm}

For a SM top quark heavier than the $W$--boson, the dominant decay
mode of the top quark is
the weak two--body decay $t \ra b W^+$.
In this mode, the top quark will analyze its own
polarization [\ref{toppol}].

QCD and SM electroweak radiative corrections to the decay width
of $t \ra b W^+$ were found to be $\sim -10$\% and $\sim 1$ \%,
respectively [\ref{cslone}].
The corrections to $\Gamma(t \ra b W^+)$, from models such as a two
Higgs doublet model and the Minimal Supersymmetric Standard Model (MSSM),
can reach the level of 1\% in favorable cases [\ref{cslonep}].

An extension of the standard Higgs sector with two Higgs doublets
has both charged and neutral Higgs bosons. If the charged
Higgs boson is lighter than the top quark, the branching ratio for the decay
$t \ra b H^+$ could be comparable to that for $t \ra b W^+$
[\ref{csltwo}].
The QCD corrections could reduce the $t \ra b H^+$ decay rate
by more than 10\% [\ref{csltwop}].
The electroweak radiative corrections to $t \ra b H^+$
were calculated and found to reduce the partial width by a few
to 10 percent, depending strongly on the parameter
$\tan \beta$ and the top quark mass [\ref{cslthree}].
The supersymmetric non-standard decay mode,
$t\ra\tilde{t_1}\widetilde{Z_1}$ has been examined
in Ref. [\ref{bdggt}].

Another interesting channel for the decay of the top quark is
the Flavor Changing Neutral Current (FCNC) decay mode.
In the SM, the branching ratios for the FCNC decay modes
were found to be too small
to be detected: $Br(t \ra cH) \sim 10^{-7}$,
$Br(t \ra cg) \sim 10^{-10}$, $Br(t \ra cZ) \sim 10^{-12}$,
$Br(t \ra c\gamma) \sim 10^{-12}$ [\ref{cslfour}].
The branching ratios of these modes
in two Higgs doublet models
or the MSSM could be enhanced by 3--4 orders of magnitudes
if one pushes the parameters far enough [\ref{cslfive},\ref{cslsix}].
It is thus a prediction of the SM and the MSSM that no large FCNC decays
exist for top quarks, so if any is detected it is beyond these approaches.
In some models, the branching ratio of the FCNC $(t \ra cH)$ decays may
be significantly enhanced, of the order $1\%$,
due to large Yukawa couplings [\ref{hou}].

\vfill
\eject

\noindent\makebox[\textwidth][l]{\bf 6. How to Detect Top Quarks}
\vspace{0.1cm}

The strategies in detecting SM top quark via the QCD processes at
the Tevatron, LHC and SSC have been studied in detail.
We refer the readers to the Proceedings of the 1990 Summer Study on High Energy
Physics [\ref{elb}], the SDC Technical Design Report [{\ref{sdc}],
the GEM Technical Design Report [\ref{gem}], and the
Proceedings of the Large Hadron Collider Workshop [\ref{lhc}].
Other results on top pair signatures can be found,
for instance, in Refs.~[\ref{verone},\ref{dalitz}],
[\ref{vertwo},\ref{dalitz}] and [\ref{verthree}]
for the dilepton, the single lepton and the all--jet modes,
respectively.

In this Workshop, H.~Baer, C.~Chen and M.~Reno  presented their
study on the jet activity associated with pair production  of
top quarks at hadron colliders [\ref{baer}].
Their approach is based upon a merger
of $2 \ra 3$ matrix elements with initial and final state
parton showers, thus yielding a calculation of $t\bar t$ plus
multi-gluon production, where the hardest gluon radiation is constrained
to agree with tree-level matrix element results.
It is found that the merged calculation yields more jet
activity in the forward region than pure matrix element estimates,
but also more jet activity in the central region.

The strategies for  observing a top quark from the
$W$--gluon fusion process at
the Tevatron are extensively discussed in Ref.~{[\ref{wgone}}].
With proper kinematic cuts, with mass of 140 GeV and
including the branching ratio $W \ra e, \, {\rm or} \, \mu$,
there should be about 10
$t$ or $\bar t$ signal events from this process produced
at the Tevatron with a 100 ${\rm pb}^{-1}$ at $\sqrt{S}=1.8$ TeV.
The major background is from $W+\, jets$ production
 which is of the same order as
the signal rate with $b$--tagging using a vertex detector. The $b$--quark
tagging efficiency is assumed to be 50\% with no misidentification.

A preliminary study shows that it should be  possible to study top quarks
produced via this process at the SSC and the LHC if a $b$-vertex detector
is used [\ref{wgthree}]. There are about 20,000
signal events and 700 background events for a 140 GeV top quark produced
via the $W$--gluon fusion process at the SSC, where the branching ratios
for $W^\pm \ra e^\pm$ or $\mu^\pm$ are included.
 This result is obtained by assuming that
it is possible to detect a jet within rapidity 4.0 and with a minimum
transverse momentum of 40 GeV.
This result of a large signal-to-background ratio is mainly
due to the characteristic features of the transverse momentum and
rapidity distributions of the spectator quark which emitted the virtual $W$
in the $W$-gluon fusion process.
However, if the rapidity coverage of the SSC/LHC detector is smaller,
then the efficiency of keeping signal events is lower and the
background event rate becomes larger [\ref{lhc}].

In models such as the MSSM and the two Higgs doublet models,
top quarks can also decay into a charged Higgs boson.
A.~Pomarol and C.~Schmidt showed in this workshop that the differences
in the lepton energy spectra from top quark decay to $H^+$ and $W^+$
may be useful in enhancing the charged Higgs signal and for
detecting the top quark when the branching ratio of
$t \ra b H^+ \ra b l^+ \nu$ is large [\ref{pomarol}].
Further analysis is required in order to optimize the
use of lepton energy information, especially in conjunction
with other techniques for enhancing  the $H^+$ signal, such as $\tau$
polarization [\ref{taupol}] or $b$--tagging [\ref{btag}].

\vspace{0.4cm}
\noindent\makebox[\textwidth][l]{\bf 7. Measuring the Top Quark Mass}
\vspace{0.1cm}

Based on the studies performed in Refs.~[\ref{sdc}],
{[\ref{gem}]} and {[\ref{lhc}]},
the top quark mass can be measured to within about 1.6\%.
The first method is to study the $e \mu$ mode of the $t \bar t$
pair [\ref{verone}].
After measuring the branching ratio of $t \ra b W^+(\ra l^+\nu)$,
one can deduce the mass of the top quark from the calculated event rate
predicted by the SM. The second method is to measure the invariant mass
of the top quark via $t \ra b W^+(\ra q_1 \bar q_2)$ [\ref{vertwo}].
The third method, which
gives the best measurement of $m_t$, is to measure the invariant mass
$M_{e\mu}$ of
$e^+$ and $\mu^-$ from the decay of $t \ra b (\ra \mu^-) W^+(\ra e^+\nu) $.

It has been shown that $M_{e\mu}$ is not sensitive to the QCD corrections
[\ref{qqttqcd}]. The distribution of $M_{e\mu}$
depends on the polarization of the top quark [\ref{steve}].
Without any kinematical cuts on the decay products of the top quark,
$M_{e\mu}$ should be the same for either a left--handed or a right--handed
top. When kinematical cuts needed to suppress the
backgrounds are applied, the mean value of $M_{e\mu}$ can be
different for a left--handed or a right--handed top [\ref{gal}].
As discussed in section 4, the top quarks produced from
the QCD processes are almost unpolarized. However, in the $W$--gluon fusion
process, they are almost one
hundred percent longitudinally polarized.
Therefore, the effects of the polarization of the top quark
should be carefully included when measuring the mass of the top quarks
produced by the $W$--gluon fusion process.
According to the studies done by G.A. Ladinsky in this workshop,
$100\%$ polarization would affect the mass measurement of the
top quark by  a few percent.

\vspace{0.4cm}
\noindent\makebox[\textwidth][l]{\bf 8. Measuring the Top Quark Width}
\vspace{0.1cm}

Reference~{[\ref{steve}]}
shows that the intrinsic width of a SM top quark
cannot be measured at the SSC and the LHC
through the usual QCD processes. For instance,
the intrinsic width of a 140 GeV SM top quark is
about 0.6 GeV and the full width at half maximum of the reconstructed
top quark invariant mass will be
about 11 GeV including the detector resolution
effects which smear the final state parton momenta. A similar conclusion was
also given from a hadron level analysis
presented in the SDC Technical Design Report;
the top quark reconstructed
invariant mass has a width of 9 GeV for a 150 GeV top quark mass [\ref{sdc}].
Can the top quark width~$\width$ be measured better than the factor
$11/0.6 \sim 20$ mentioned above? The answer is yes.  As
pointed out in Ref.~{[\ref{steve}]},
the width $\width$ can be measured by counting the production rate of top
quarks
from the $W$--$b$ fusion process. This is {\it equivalent} to the $W$--gluon
fusion process with the proper treatment of the bottom
quark and the $W$ boson as partons inside the
hadron.
 The $W$--boson which interacts with the $b$--quark to produce the top
quark can be treated as an on--shell boson
in the leading log approximation.
 The moral is that even
with the approximations considered,
a factor of 2 in the uncertainty for the production
rate for this process gives a
factor of 2 in the uncertainty for the measurement of $\width$.
This is still much more accurate than what can be done
at the SSC and the LHC through the usual QCD processes.
Therefore, this is an extremely important measurement to make at
the Tevatron because it directly tests the SM coupling of $t$-$b$-$W$.
Similarly, this production mechanism is also useful
at the SSC/LHC for detectors covering a large enough rapidity region,
as discussed in section 6.

After the top quark is found, one can
measure the branching ratio of $t \ra b W^+(\ra l^+\nu)$ by the
ratio of $(2l+\,jets)$ and $(1l+\,jets)$
event rates from $t \bar t$ production.
Then, a model independent measurement of the decay
width~$\width$ can be made by counting the production rate
of $t$ in the $W$--gluon fusion process.
Should the top quark be found to be different from the SM expectations,
we would then look for
non-standard decay modes of the top quark [\ref{kaner}].
It is still important to measure at least one
partial width~$\width$ precisely, in order to discriminate between
different models of new physics.

\vspace{0.4cm}
\noindent\makebox[\textwidth][l]{\bf 9. CP Violation in Top Quarks}
\vspace{0.1cm}

If CP is violated in the production of top quarks at the Tevatron,
the production rate of $t$ from
 $ \bar p p(W^+ g) \ra t \bar b X$
would be different from that of $\bar t$ from
$\bar p p(W^- g) \ra \bar t  b X$.
Therefore, one can detect large CP violation effects
by observing the difference in the production rates of $t$ and $\bar t$.
Large CP violating effects are required to have the cosmological
baryon asymmetry produced at the weak phase transition [\ref{baryon}].

In the $W$--gluon fusion process, the top quark is
almost one hundred percent longitudinally
polarized. This allows us to probe CP violation
in the decay process
$t \ra W^+ b \ra l^+ \nu_l b$.
The most obvious observable for this purpose
is
the expectation value of the time--reversal quantity
$
\vec{\bf\sigma_t} \cdot (\widehat{\bf p}_b \times \widehat{\bf p}_{l})
$,
where $\vec{\bf\sigma_t}$ is the polarization vector of $t$, and
$\widehat{\bf p}_b$ ($\widehat{\bf p}_l$) is the unit vector of the
$b$ ($l^+$) momentum in the rest frame of the top quark [\ref{nucl}].
This was suggested in Ref.~{[\ref{toppol}]} and further
studied in Ref.~{[\ref{gunion}]}.

Furthermore, it was shown in Ref.~{[\ref{peskin}]} that it is possible
to study CP violation effects, via the usual QCD processes,
by observing the asymmetry in the energies
of the two leptons from the decay of the $t \bar t$ pairs
produced at the SSC/LHC.
A detailed study, including all the possible interactions for
top quarks,
on how to measure the CP violation effects at the SSC and LHC
colliders is given in Ref.~{[\ref{imkane}]} by C.~Im and G.L.~Kane.

\vspace{0.4cm}
\noindent\makebox[\textwidth][l]{\bf 10. Top Quark Couplings }
\vspace{0.1cm}

Recently, there has been a lot of interesting work studying
the effects of large Yukawa couplings due to heavy top quarks
in the Supersymmetric Grand Unified Theories (SUSY--GUT).
Some of these effects, such as the unification of the gauge couplings,
the evolution of the Yukawa couplings,
and the evolution of the CKM matrix, etc.,
 were discussed by V. Barger in this Workshop [\ref{barger}].

In the SM, the top quark gains mass through the Yukawa coupling
to the Higgs boson $H$. To test the dynamics of the generation of the
fermion mass, we should also
measure the coupling of the $Ht \bar t$. The direct measurement of this
coupling can be done by studying the production
of top quarks in $gg, q \bar q \ra t \bar t H$ and
$qg,  {\bar q} g \ra t \bar b H$ [\ref{ttH},\ref{tbH}].
Since the signal rate is not large at the SSC/LHC, it might be difficult to
get a precise measurement from these processes.
If the luminosity of the collider is high enough to produce a
large number of $qg, {\bar q} g \ra t \bar b H$ events, one can
test the relative sign of the $Ht \bar t$ and $H W^+ W^-$ couplings
which depend on the underlying dynamics of symmetry breaking.
Another place to look for the
coupling of $H t \bar t$
is in the production of the Higgs boson from the gluon fusion
process through a top quark triangle loop. The QCD next-to-leading-order
calculation of this cross section was performed
a several years ago [\ref{qcdh}].
To get a precise measurement of the $H t \bar t$ coupling through
radiative corrections to Higgs boson production, we have to know
the kinematics of the Higgs boson after including the multiple gluon
emission effects from initial state radiation. This calculation of the
QCD-gluon resummation was performed in Ref.~{[\ref{resum}]}.

\vspace{0.4cm}
\noindent\makebox[\textwidth][l]{\bf 11. Top Quark Decays as Backgrounds }
\vspace{0.1cm}

Up to now, we have only discussed the top quark as a signal;
it is also
one of the most important backgrounds in probing new physics
at high energy colliders.
For example, Refs.~{[\ref{wwone}]} and {[\ref{wwtwo}]}
show how to suppress the backgrounds
associated with the top quark in order
to study the strongly interacting
longitudinal $W$-boson system in the TeV region to probe the mechanism of the
electroweak symmetry breaking.
These top quark background estimates depend on a forward jet tagging
requirement. This is precisely the region where multiple gluon emission effects
can be large. Ref. [\ref{baer}] estimates these effects, and finds the multiple
gluon effects lead to larger $t\bar t$ backgrounds if only forward jet tagging
is required. However, a combination of forward jet tag and central jet veto
yields results perhaps coincidentally close to $2\ra 3$ calculations, due
to extra gluon radiation in the central region as well.
Top quarks as backgrounds for some other
processes were discussed in Ref.~{[\ref{www}]}.

In conclusion, we have to study the top quark and its production
in detail to test the
SM, to probe directly for new physics and to allow other probes for new
physics. Most of these goals can be accomplished at the Tevatron
and  SSC/LHC colliders.

\vglue 0.5cm
{\elevenbf \noindent Acknowledgements \hfil}
\vglue 0.5cm

We would like to thank
those who participated in the discussion section of the
Heavy Flavor Physics group at the workshops.  Special thanks are due to
N.G. Deshpande, R. Hollebeek and M. Doncheski.
This work was supported in part by the Michigan State University,
Texas National Research Laboratory Commission grant RGFY9240, and the
U. S. Department of Energy, Division of High Energy Physics, contract number
W-31-109-ENG-38.

\begin{reflist}

\item \label{china}
C.--P. Yuan, preprint MSUHEP-93/10, July 1993,
to appear in the {\it Proceedings of the Workshop on
Particle Physics at the Fermi Scale},
Beijing, China, May 27 -- June 4, 1993.

\item \label{nmshaw}
N. M. Shaw, talk presented at this Workshop.

\item \label{diehl}
T. Diehl, talk presented at this Workshop.

\item \label{slwu}
S. L. Wu, talk presented at this Workshop.

\item \label{toppol}
G.L.~Kane, G.A.~Ladinsky and C.-P.~Yuan, {\it Phys.~Rev.}
{\bf D45} (1992) 124.

\item \label{decay}
I.I.Y. Bigi, Yu L. Dokshitzer, V.A. Khoze, J.H. Kuhn, and P. Zerwas,
 {\it Phys.~Lett.} {\bf B181} (1986) 157;
L.H. Orr and J.L. Rosner, {\it Phys.~Lett.} {\bf B246} (1990) 221;
{\bf 248B} (1990) 474(E).

\item \label{nloqcd}
P. Nason, S. Dawson and R.K. Ellis, \jourtpol
Nucl.~Phys. &{B303}, 607 (1988); {\bf B327}, 49 (1989);
W. Beenakker, H. Kuijf, W.L. van Neerven and J. Smith,
\jourtpol Phys.~Rev. &{D40}, 54 (1989);
R. Meng, G.A. Schuler, J. Smith and W.L. van Neerven,
\jourtpol Nucl.~Phys. &{B339}, 325 (1990).

\item \label{qcdtpol}
F.~Berends et al., in the report of the Top Physics Working
Group from the {\it Proceedings of the Large Hadron
Collider Workshop}, 4-9 October 1990, Aachen, ed. G.~Jarlskog and D.~Rein,
CERN publication CERN 90-10;
A.~Stange and S.~Willenbrock, Fermilab preprint
FERMILAB-PUB-93-027-T, Feb. 1993;
W. Beenakker, A. Denner, W. Hollik, R. Mertig, T. Sack and
D. Wackeroth, preprint MPI-Ph/93-20, May 1993.

\item \label{qqttqcd}
 C. Kao, G. Ladinsky, and C.--P. Yuan, preprint
MSUHEP 93/04, 1993.

\item \label{wgone}
C.--P. Yuan, \jourtpol Phys.~Rev. &{D41}, 42 (1990);
D. Carlson and C.--P. Yuan, {\it Phys. Lett.} {\bf B306} (1993) 386.

\item \label{wgtwo}
S. Dawson, \jourtpol Nucl.~Phys. &{B249}, 42 (1985);
S. Dawson and S. S. D. Willenbrock, \jourtpol Nucl.~Phys. &{B284}, 449 (1987);
S. S. D. Willenbrock and D. A. Dicus, \jourtpol Phys.~Rev. &{D34}, 155 (1986);
F. Anselmo, B. van Eijk and G. Bordes, \jourtpol Phys.~Rev. &{D45}, 2312
(1992);
T. Moers, R. Priem, D. Rein and H. Reithler in {\it Proceedings of Large
Hadron Collider Workshop}, preprint CERN 90-10, 1990;
R.K. Ellis and S. Parke, \jourtpol Phys.~Rev. &{D46}, 3785 (1992).

\item \label{goldstein}
W.G.D.~Dharmaratna and G.R.~Goldstein,
{\it Phys.~Rev.} {\bf D41} (1990) 1731.

\item \label{trans}
G.L. Kane, J. Pumplin and W. Repko, {\it Phys.~Rev.~Lett.} {\bf 41}
(1978) 1689;
A. Devoto, G.L. Kane, J. Pumplin and W. Repko,
{\it Phys.~Lett.} {\bf B90} (1980) 436.

\item \label{cslone}
J.~Liu and Y.--P. Yao, University of Michigan preprint UM-TH-90-09;
C.S. Li, R.J. Oakes and T.C. Yuan, {\it Phys. Rev.} {\bf D43} (1991) 3759;
M. Jezabek and
J.H. Kuhn, {\it Nucl.~Phys.} {\bf B314} (1989) 1;
{\it ibid} {\bf B320} (1989) 20;
A. Denner and T. Sack, {\it Nucl. Phys.} {\bf B358} 46 (1991);
Z. Phys. {\bf C46},
653 (1990);
G. Eilam, R. R. Mendel, R. Migneron,
and A. Soni, {\it Phys. Rev. Lett.} {\bf 66} (1991) 3105;
B. A. Irwin, B. Margolis, and H. D. Trottier,
{\it Phys. Lett.} {\bf B256}, 533 (1991);
J. Liu and Y.--P. Yao,
{\it Int.J.Mod.Phys.} {\bf A6} (1991) 4925;
T.C. Yuan and C.--P. Yuan, {\it Phys. Rev.} {\bf D44} (1991) 3603.

\item \label{cslonep}
B. Grzadkowski, W. Hollik, {\it Nucl. Phys.} {\bf B384} (1992) 101;
J.M. Yang and C.S. Li, preprint CQU-TH-93-1, 1993;
C.S. Li, J.M. Yang and B.Q. Hu, preprint CQU-TH-92-16, 1992.

\item \label{csltwo}
For a review, see J.F. Gunion, H.E. Haber, G.L. Kane and S. Dawson,
{\it The Higgs Hunter's Guide}, (Addison--Wesley, Menlo Park, 1990).

\item \label{csltwop}
C.S. Li and T.C. Yuan, {\it Phys. Rev.} {\bf D42} (1990) 3088;
A. Mendez and A. Pomarol, {\it Phys. Lett.} {\bf B252} (1990) 461;
C.S. Li,  Y.S. Wei and J.M. Yang, {\it phys. Lett.} {\bf B285} (1992) 137;
J. Liu and Y.--P. Yao, {\it Phys. Rev.} {\bf D46} (1992) 5196;
A. Czarnecki and S. Davidson, {\it Phys. Rev.} {\bf D47} (1993) 3063.

\item \label{cslthree}
A. Mendez and A. Pomarol, {\it phys. Lett.} {\bf B265} (1991) 177;
C.S. Li, B.Q. Hu and J.M. Yang, {\it phys. Rev.} {\bf D47} (1993) 2866;
J.M. Yang, C.S. Li and B.Q. Hu, {\it phys. Rev.} {\bf D47} (1993) 2872.

\item \label{bdggt}
H. Baer, M. Drees, R. Godbole, J. Gunion and X. Tata, Phys. Rev. {\bf D44}
(1991) 725.

\item \label{cslfour}
W. Buchmuller and M. Gronau, {\it Phys. Lett.} {\bf B220} (1989) 641;
H. Fritzsch, {\it Phys. Lett.} {\bf B224} (1989) 423;
B. Dutta-Roy, B.A. Irwin, B. Margolis, J. Robinson, H.D. Trottier,
C. Hamazaoui, {\it Phys. Rev. Lett.} {\bf 65} (1990) 827;
J.L. Diaz-Cruz, R. Martinez, M.A. Perez, A. Rosado, {\it Phys. Rev.} {\bf D41}
(1990) 891.

\item \label{cslfive}
G. Eilam, J.L. Hewett and A. Soni, {\it Phys. Rev.} {\bf D44} (1991) 1473;
B. Grzadkowski, J.F. Gunion and P. Krawczyk,
{\it Phys. Lett.} {\bf B268} (1991) 106;
C.S. Li, R.J. Oakes and J.M. Yang, preprint CQU-TH-93-2, 1993.

\item \label{cslsix}
J.M. Yang and C.S. Li, preprint CQU-TH-93-5, 1993.

\item \label{hou}
W. S. Hou, {\it Phys. Lett.} {\bf B296} (1992) 179.

\item \label{elb}
Research Directions for the Decade,
{\it Proceedings of the 1990 Summer Study on High Energy
Physics}, June 25 -- July 13, 1990, Snowmass, Colorado, edited by E.L. Berger.

\item \label{sdc}
 SDC Technical Design Report, preprint SDC-92-201, 1992.

\item \label{gem}
GEM Technical Design Report, preprint GEM-TN-93-262, April 1993.

\item \label{lhc}
Top Physics Working
Group from the {\it Proceedings of the Large Hadron
Collider Workshop}, 4-9 October 1990, Aachen, ed. G.~Jarlskog and D.~Rein,
CERN publication CERN 90-10.

\item \label{verone}
 H. Baer, V. Barger, J. Ohnemus, and R.J.N. Phillips,
{\it  Phys. Rev.} {\bf D42} (1990) 54 and references therein;
 T. Han and S. Parke, preprint FERMILAB-PUB-93/105-T.

\item \label{dalitz}
R.H. Dalitz and G.R. Goldstein, preprint OUTP-93-16P, 1993.

\item \label{vertwo}
H. Baer, V. Barger, and R.J.N.~Phillips,
{\it Phys. Rev.} {\bf D39} (1989) 3310;
F.A. Berends, H. Kuif, B. Tausk, and W.T. Giele,
{\it Nucl. Phys.} {\bf B357} (1991) 32;
F.A. Berends, J.B. Tausk, and W.T. Giele,
{\it Phys. Rev.} {\bf D47}  (1993) 2746;
V. Barger, J. Ohnemus, and R.J.N. Phillips,
University of Wisconsin report MAD/PH/777.

\item \label{verthree}
J.M. Benlloch, N. Wainer, and W.T. Giele,
preprint FERMILAB-PUB-93/060-T.

\item \label{baer}
H. Baer, Chih-hao Chen and M.H. Reno, preprint FSU-HEP-930625,
{\it Phys. Rev.} {\bf D.} (in press) (1993).

\item \label{wgthree}
 D. Carlson and C.--P. Yuan, in preparation.

\item \label{pomarol}
A. Pomarol and C. Schmidt, talk given at this Workshop.

\item \label{taupol}
B.K. Bullock, K. Hagiwara and A.D. Martin,
{\it Phys. Lett.} {\bf B273} (1991) 501;
{\it Phys. Rev. Lett.} {\bf 67} (1991) 3055;
D.P. Roy, {\it Phys. Lett.} {\bf B277} (1992) 183.

\item \label{btag}
R.M. Barnett, R. Cruz, J.F. Gunion and B. Hubbard,
{\it Phys. Rev.} {\bf D47} (1993) 1048.

\item \label{steve}
 S. Mrenna and C.--P. Yuan, {\it Phys. Rev.} {\bf D46} (1992) 1007.

\item \label{gal}
G.A. Ladinsky, preprint MSUHEP-93/11, August 1993.

\item \label{kaner}
For a review, see, G.L. Kane, preprint UM-TH-91-32, 1991,
published in Mexico City High Energy Phenomenology (1991) 241.

\item \label{baryon}
N.~Turok and J.~Zadrozny {\journal Phys. Rev.&65 (90) 2331};
M.~Dine, P.~Huet, R. Singleton and L.~Susskind
{\journal Phys. Lett.&B257 (91) 351};
L.~McLerran, M. Shaposhnikov, N.~Turok and
M.~Voloshin{\journal Phys. Lett.&B256 (91) 451};
A.~Cohen, D.~Kaplan and A.~Nelson{\journal Phys. Lett.&B263 (91) 86}.

\item \label{nucl}
J.D. Jackson, S.B. Treiman and H.W. Wyld, Jr., \jourtpol
Phys.~Rev. & {106} (1957) 517;
R.B. Curtis and R.R. Lewis, \jourtpol Phys.~Rev. & {107} (1957) 543.

\item \label{gunion}
B. Grzadkowski and J.F. Gunion, preprint UCD-92-7, 1992.

\item \label{peskin}
C.R.~Schmidt and M.E.~Peskin, {\it Phys. Rev. Lett.} {\bf 69} (1992) 410.

\item \label{imkane} C. Im, G. Kane and P. Malde, preprint UM-TH-92-27,
to appear in {\it Phys. Lett.} {\bf B.} (1993).

\item \label{barger}
V. Barger, M.S. Berger, P. Ohmann and R.J.N. Phillips,  preprint MAD/PH/781,
\break
August 1993.

\item \label{ttH}
 W. Marciano and F. Paige, {\it Phys.~Rev.~Lett.} {\bf 66} (1991) 2433;
J. Gunion, {\it Phys.~Lett.} {\bf B261} (1991) 510;
R. Kleiss, Z. Kunszt and W. Stirling,
\jourtpol Phys.~Lett. &{B253} (1991) 269.

\item \label{tbH}
 J. Diaz-Cruz and O. Sampayo,
\jourtpol Phys.~Lett. &{B276} (1992) 211;
W. Stirling and D. Summers, \jourtpol Phys.~Lett. &{B283} (1992) 411.

\item \label{qcdh}
S. Dawson, {\it Nucl. Phys.} {\bf B359} (1991) 283;
A. Djouadi, M. Spira and P.~M.~Zerwas, {\it Phys. Lett.} {\bf B264}
(1991) 440.

\item \label{resum}
P. Agrawal, J. Qiu and C.--P. Yuan,
in preparation; C.--P. Yuan, {\it Phys. Lett.} {\bf B283} (1992) 395;
R. P. Kauffman, {\it Phys. Rev.} {\bf D45} (1992) 2273.

\item \label{wwone}
G. Ladinsky and C.--P. Yuan, {\it Phys. Rev.} {\bf D43} (1991) 789;
R. Kauffman and C.--P. Yuan, {\it Phys. Rev.} {\bf D42} 42 (1990) 956.

\item \label{wwtwo}
 J. Bagger, V. Barger, K. Cheung, J. Gunion,
T. Han, G. Ladinsky, R. Rosenfeld
and C.--P. Yuan, preprint MSUHEP-93/05, 1993;
 and the references therein.

\item \label{www}
V. Barger, A.L. Stange and R. Phillips, {\it Phys. Rev.} {\bf D45} (1992) 1484.

\end{reflist}
\newpage

\end{document}